\documentclass[twocolumn,secnumarabic,amssymb, nobibnotes, aps, prb,groupedaddress,superscriptaddress]{revtex4-2}
\usepackage{amsmath,graphicx,latexsym,times,color}
\usepackage{setspace}
\usepackage{hyperref}
\usepackage{array}
\usepackage{titlesec}
\usepackage{physics}
\usepackage{gensymb}
\usepackage{nccmath}
\usepackage{empheq} 
\usepackage[cmyk,dvipsnames]{xcolor}

\newcommand{\V}[1]{\ensuremath{\mathbf{#1}}} 

\let\oldtimes\times  
\renewcommand\times{{\oldtimes}}

\renewcommand{\vec}[1]{\mathbf{#1}}

\definecolor{dsgrey}{rgb}{0,1,1}

\begin{document}


\title{Emergence of multiple topological spin textures in an all-magnetic van der Waals heterostructure}

	\affiliation{Universit\'e de Toulouse, CNRS, CEMES, Toulouse, France}
     \affiliation{Institute of Theoretical Physics and Astrophysics, University of Kiel, Leibnizstrasse 15, 24098 Kiel, Germany}  
     \affiliation{Science Institute and Faculty of Physical Sciences, University of Iceland, VR-III, 107 Reykjavík, Iceland} 
     \affiliation{Kiel Nano, Surface, and Interface Science (KiNSIS), University of Kiel, 24118 Kiel, Germany} 
 
	\author{Moritz A. Goerzen}
	\affiliation{Universit\'e de Toulouse, CNRS, CEMES, Toulouse, France}
    \affiliation{Institute of Theoretical Physics and Astrophysics, University of Kiel, Leibnizstrasse 15, 24098 Kiel, Germany}

 	\author{Tim Drevelow}
	\affiliation{Institute of Theoretical Physics and Astrophysics, University of Kiel, Leibnizstrasse 15, 24098 Kiel, Germany}

    \author{Hendrik Schrautzer}
    \affiliation{Institute of Theoretical Physics and Astrophysics, University of Kiel, Leibnizstrasse 15, 24098 Kiel, Germany}
    \affiliation{Science Institute and Faculty of Physical Sciences, University of Iceland, VR-III, 107 Reykjavík, Iceland}

	\author{Soumyajyoti Haldar}
	\affiliation{Institute of Theoretical Physics and Astrophysics, University of Kiel, Leibnizstrasse 15, 24098 Kiel, Germany}

	\author{Stefan Heinze}
	\affiliation{Institute of Theoretical Physics and Astrophysics, University of Kiel, Leibnizstrasse 15, 24098 Kiel, Germany}
	\affiliation{Kiel Nano, Surface, and Interface Science (KiNSIS), University of Kiel, 24118 Kiel, Germany}

\author{Dongzhe Li}
\email[Contact author: ]{dongzhe.li@cemes.fr}
	\affiliation{Universit\'e de Toulouse, CNRS, CEMES, Toulouse, France}

	\date{\today}
	
\begin{abstract}
{Magnetic solitons such as skyrmions and bimerons show great promise for both fundamental research and spintronic applications. Stabilizing and controlling topological spin textures in atomically thin van der Waals (vdW) materials has gained tremendous attention due to high tunability, enhanced functionality, and miniaturization. Here, we present an efficient spin-spiral approach based on first-principles, a method for mapping magnetic interactions from collective models onto arbitrary lattice symmetries, such as hexagonal and honeycomb lattices. Using atomistic spin models parametrized from first-principles, we predict the emergence of multiple topological spin textures in an all-magnetic vdW heterostructure Fe$_3$GeTe$_2$/Cr$_2$Ge$_2$Te$_6$ (FGT/CGT) -- an experimentally feasible system. Interestingly, the FGT layer favors out-of-plane magnetization, whereas the CGT layer prefers in-plane magnetocrystalline anisotropy. Néel-type nanoscale skyrmions are formed at zero field in the FGT layer due to interfacial Dzyaloshinskii-Moriya interaction (DMI), while nanoscale bimerons and antibimerons can co-exist in the CGT layer by the interplay between exchange frustration and DMI. Using the collective approach we apply, we reveal significant discretization effects in hexagonal and honeycomb geometries. In particular, we demonstrate that the lifting of geometric exchange frustration on the honeycomb significantly affects soliton barriers and pinning energetics. These fundamental results not only highlight the importance of spin simulations in discrete models for topological magnetism, especially in 2D materials, but may also help to pave the way for solitonic devices based on atomically thin vdW heterostructures.
}
\end{abstract}
	
	\maketitle

\section{INTRODUCTION}

Magnetic skyrmions, vortex-like localized spin structures with nontrivial topology, have attracted considerable attention because of their rich physics and potential applications in future spintronic devices \cite{fert2017magnetic,gobel2021beyond,Psaroudaki2021}. Unique properties of skyrmions for applications include their nanoscale size \cite{romming2013writing,meyer2019isolated}, good thermal stability due to their integer topological charge \cite{gobel2021beyond}, manipulation by electric currents \cite{fert2013,sampaio2013nucleation}, full-electrical detection \cite{hanneken2015electrical,maccariello2018electrical,Perini2019,Dongzhe2023,chen2024all}, and ultrafast topological switching \cite{buttner2021observation,dabrowski2022all,khela2023laser}. In analogy to magnetic skyrmions in perpendicularly magnetized systems, bimerons are the combination of a meron and an anti-meron and have been interpreted as in-plane magnetized versions of magnetic skyrmions \cite{bimeron_PRB2019}. We recently demonstrated that bimerons cannot be regarded as in-plane counterparts of skyrmions because of their distinct structural symmetry \cite{Moritz_bimeron2025}. Due to the in-plane magnetization background, bimerons offer additional degrees of freedom compared to conventional skyrmions. Materials hosting multiple topological spin textures add richness to the field, allowing the design of diverse spintronics building blocks. 

The demand for device miniaturization in modern electronics pushes the skyrmion playground from conventional transition-metal multilayers to two-dimensional (2D) van der Waals (vdW) materials consisting of weakly bonded 2D layers. Between 2019 and 2020, two independent experimental groups reported the first observation of skyrmions in the vdW magnets Cr$_2$Ge$_2$Te$_6$ \cite{han2019topological} and Fe$_3$GeTe$_2$ \cite{ding2019observation}. Later on, several experimental groups observed various skyrmion lattices in 2D magnets vdW heterostructures \cite{wu2020neel,Park2021,wu2022van,powalla2023seeding}. One key parameter for generating skyrmions is the relativistic Dzyaloshinskii-Moriya interaction (DMI), which originates from spin-orbit coupling (SOC) and relies on broken inversion symmetry. From the theoretical point of view, in the past years, in order to obtain large DMI in 2D materials, a comprehensive material survey has been performed in Janus layers which exhibit intrinsic inversion asymmetry \cite{Liang2020,Yuan2020,Changsong2020,du2022spontaneous,Megha_2025}, using proximity in vdW heterostructures \cite{sun2020controlling,sun2021manipulation,yang2020creation,Dongzhe2022_fgt,Dongzhe_prb2023}, decorating light atoms (such as oxygen or lithium atoms) on the surface of 2D magnets or doping 2D magnets \cite{Park2021,zhang2022room,Weiyi2024,fgt-Li}, and in ferroelectric structures \cite{Chao-Kai_2021,Huang2022,Peixuan_2024}. 

\begin{figure*}[tbp]
	\centering
	\includegraphics[width=1.0\linewidth]{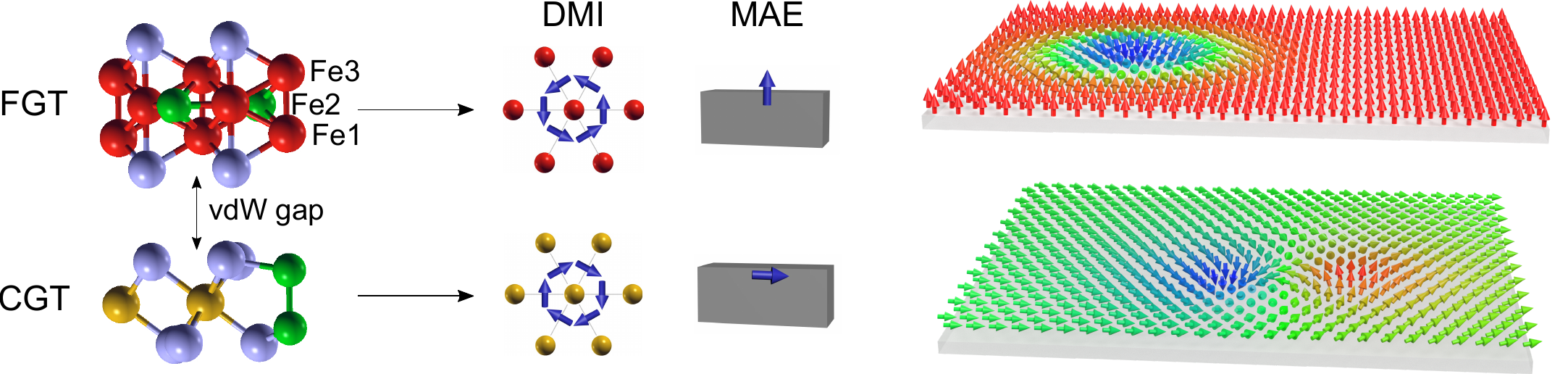}
	\caption{\label{cgt_fgt_stucture} Crystal structure of the Fe$_3$GeTe$_2$/Cr$_2$Ge$_2$Te$_6$ (FGT/CGT) van der Waals interface. Side view of the atomic structure of FGT/CGT. The induced DMI vectors (top view) and easy-axis for the FGT and CGT layers at the interface (side view) are shown schematically. On the right, we show the topological solitons stabilized in the FGT/CGT heterostructure: skyrmions form spontaneously in the FGT layer, while bimerons appear in the CGT layer.
 }
\end{figure*}

For practical applications, it is crucial to realize the creation, transformation, and manipulation of skyrmions and bimerons via external stimuli. The skyrmion–bimeron phase transition has been demonstrated in experiments using magnetic field \cite{yu2018transformation}, temperature \cite{li2021field}, or electric field \cite{ohara2022reversible}. On the other hand, several theoretical works 
reported various topological spin textures (i.e., skyrmions, bimeron, vortices) in 2D vdW heterostructures \cite{sun2021manipulation,augustin2021properties,he2022multiple}. However, the studies mentioned above only focused on forming metastable skyrmions or bimerons in various ferromagnetic/non-magnetic interfaces. Fundamental properties of skyrmions and bimerons, such as collapse mechanisms and energy barriers that are crucial for future device applications, still remain largely unexplored in 2D vdW magnets and heterostructures.

Here, employing rigorous first-principles calculations and atomistic spin simulations, we predict the formation of multiple topological spin textures in an all-magnetic vdW heterostructure -- Fe$_3$GeTe$_2$/Cr$_2$Ge$_2$Te$_6$ (FGT/CGT) -- an interface that was recently synthesized experimentally \cite{wu2022van,Wang2023_fgtcgt}. By exploiting the heterostructure of two vdW ferromagnets, we stabilize two types of topological solitons at zero field: skyrmions in the FGT layer and bimerons in the CGT layer, the latter undergoing a bimeron–skyrmion transformation under an applied magnetic
field. We analyze soliton size, energy barriers, collapse mechanisms, and the role of lattice symmetry in great detail. We show that skyrmions are more strongly pinned in the honeycomb lattice than in the hexagonal lattice.

This paper is organized as follows. In Sec .~\ref {method_cal}, we briefly discuss the theoretical methods and computational details used in this work. In particular, we introduce the methodologies for parameterizing the Heisenberg model in arbitrary lattices using collective models obtained from density functional theory (DFT). In Sec.~\ref{results}, using atomistic spin models parameterized by first-principles, we present the properties of magnetic skyrmions and bimerons in FGT/CGT, including soliton size, energy barriers, collapse mechanisms, and the influence of lattice symmetry on soliton properties. We argue that lattice symmetry plays a crucial role in determining soliton energy barriers and pinning energetics. Finally, in Sec.~\ref{conclusions}, we summarize our conclusions.

\begin{figure*}[t]
    \centering
    \includegraphics[width=0.95\linewidth]{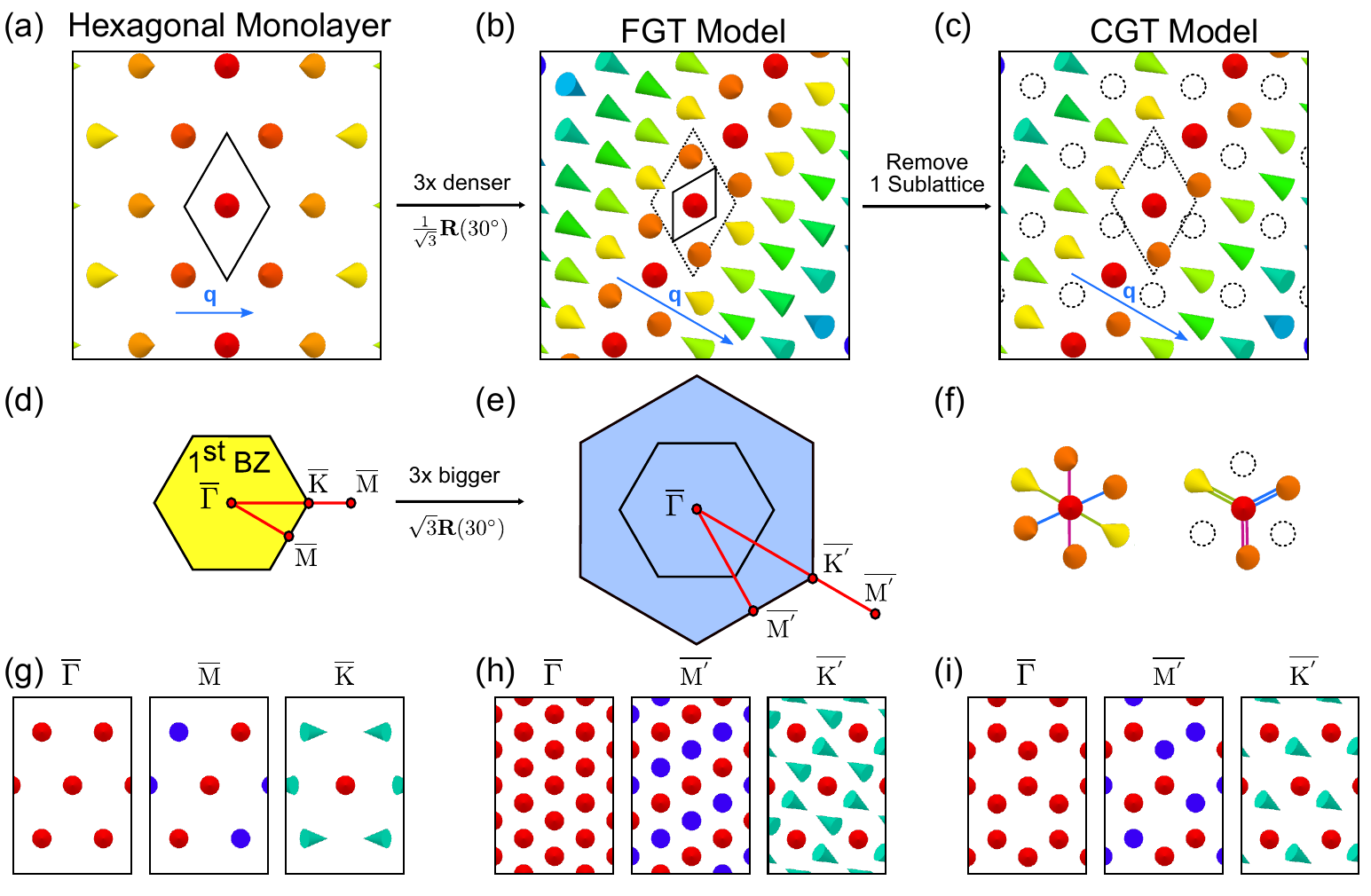}
    \caption{
    Models for fitting magnetic interactions in FGT and CGT. 
(a) Spin spiral in a magnetic monolayer model with the spin spiral vector $\mathbf{q}$.
(b) Rotated and scaled dense hexagonal layer model, which has 3 magnetic atoms in the chemical unit cell from top view, with the transformed spin spiral vector, which is used as the model for the FGT layer.
(c) Honeycomb lattice model (2 magnetic atoms per unit cell) with vacancies that would transform it into a hexagonal layer if they were filled. The honeycomb lattice models the CGT layer.  
(d) Symmetry zone of the spin spiral vector $\mathbf{q}$. It coincides with the Brillouin zone (BZ) in the case of a magnetic monolayer. The path around the irreducible wedge of the symmetry zone is shown in red.
(e) Spin spiral symmetry zone for the denser hexagonal and honeycomb lattice models. It is 3 times bigger than the BZ and slightly rotated. 
(f) Illustration of the interactions between nearest neighbors in the dense hexagonal and the honeycomb lattice model. On the hexagonal lattice, there are two nearest neighbors for each of the three high symmetry directions (green, blue, and red). In the honeycomb model, there are only half as many interactions. If they had double strength in the honeycomb model, both models would produce the same spin spiral energy dispersions.
(g)-(i) Show the states at the high-symmetry points of the different models. For the latter two, the larger BZ leads to new high-symmetry points.
}
    \label{fig:CGT-lattices}
\end{figure*}

\section{METHODS AND COMPUTATIONAL DETAILS}
\label{method_cal}

\subsection*{A. Spin model}
To investigate magnetic interactions of FGT/CGT, we map our DFT total energies to the following spin Hamiltonian
\begin{equation}
\label{spin_model}
\begin{split}
H & =-\sum_{ij}J_{ij}(\V{m}_i \cdot \V{m}_j)-\sum_{ij}\V{D}_{ij} \cdot(\V{m}_i \times \V{m}_j) \\
&  - K \sum_i (m_i^z)^2 - \mu\sum_i\V{m}_i\cdot\V{B}~,
\end{split}
\end{equation}
where $\V{m}_i$ and $\V{m}_j$ denote the normalized magnetic moments at positions $\V{R}_i$ and $\V{R}_j$, respectively, and $\mu$ represents the magnitude of the magnetic moment at each site. The four magnetic interaction terms correspond to the Heisenberg isotropic exchange, the Dzyaloshinskii–Moriya interaction (DMI), the magnetocrystalline anisotropy energy (MAE) with uniaxial anisotropy, and the Zeeman interaction. These are characterized by the parameters $J_{ij}$, $\V{D}_{ij}$, $K$, and $\V{B}$, respectively.

\subsection*{B. First-principles calculations}

For atomic relaxations, we employed DFT in the generalized gradient approximation (GGA) using the \textsc{Quantum Espresso} \cite{giannozzi2009quantum} code. A cutoff of 45 Ry was used for the wavefunctions, while a 450 Ry cutoff was employed for the charge density. We also took into account van der Waals interactions using semi-empirical dispersion corrections (DFT-D3) as formulated by Grimme \cite{grimme2010consistent}. The FGT/CGT interface is modeled using a $(1 \times 1)$ CGT unit cell matched to a $(\sqrt{3} \times \sqrt{3})$ FGT cell, resulting in a lattice mismatch of less than 0.15\%. This optimized geometry is in good agreement with Ref.~\cite{Wang2023_fgtcgt}. The Brillouin zone (BZ) has been discretized by using a $(12 \times 12 \times 1)$ $\V{k}$-points mesh. In addition, to avoid unphysical interactions in the $z$ direction (perpendicular to the surface), a vacuum space of about 25 \AA~is used in the $z$ direction.

To calculate the magnetic interaction constants in Eq.~(\ref{spin_model}), we performed DFT total-energy calculations for various collinear and noncollinear spin structures, using the full-potential linearized augmented plane-wave (FLAPW) method as implemented in the \textsc{Fleur} code \cite{fleurv26,fleurCode}. For the exchange–correlation functional, we employed the local density approximation (LDA) in the parametrization of Vosko, Wilk, and Nusair \cite{vosko1980accurate}. We applied LDA+$U$ with $U_{\text{eff}} = 0.5$ eV for CGT, following Ref. \cite{gong2017discovery}, but not for FGT, since LDA yields a magnetic moment for the Fe atom that is closer to the experimental value, as previously noted in Refs. \cite{Zhuang2016,deng2018gate}. For more computational details, see our previous work \cite{Moritz_bimeron2025}.

In short, the magnetic interactions in Eq.~(\ref{spin_model}) are determined in three steps by first-principles: (i) We determined the exchange interaction from the energy dispersions of flat spin spirals calculated without SOC, using the generalized Bloch theorem \cite{Kurz2004}. (ii) The DMI was calculated by adding the SOC contribution to the spin-spiral energy dispersion within first-order perturbation theory \cite{Heide2009,Zimmermann2014}. (iii) For the MAE, we employed the magnetic force theorem, evaluating the band-energy difference in a non-self-consistent manner, $K = E^{\text{band}}_{\perp}-E^{\text{band}}_{\parallel}$. 

\subsection*{C. Parametrization of the Heisenberg model}
\label{fitt_largeQ}

To model realistic systems, our chosen system is the 2D vdW heterostructure FGT/CGT, where a monolayer of FGT is deposited on a monolayer of CGT (see Fig.~\ref{cgt_fgt_stucture}). The in-plane components of the DMI vectors, as obtained from DFT, are also shown
in Fig.~\ref{cgt_fgt_stucture}: counterclockwise (CCW) for the FGT layer and clockwise (CW) for the CGT layer. Moreover, the MAE favors out-of-plane magnetization in the FGT layer and in-plane magnetization in the CGT layer. These particular magnetic interactions give rise to the possibility of having multiple topological 
magnetic states in one system, as we will demonstrate in the following: skyrmions are formed in the FGT layer while bimerons are stabilized in the CGT layer (see Fig.~\ref{cgt_fgt_stucture}). The two magnetic layers are coupled by van der Waals interactions. The FGT/CGT supercell contains 11 magnetic atoms, including 9 Fe atoms and 2 Cr atoms. This presents a complicated problem for calculating magnetic interactions based on spin spirals, as the FGT layer is a multilayer system (3 Fe layers), whereas the CGT layer forms a honeycomb lattice.

In this subsection, we discuss how to map DFT total energies onto the spin model. Both FGT and CGT have a complex geometric structure, which gives rise to a large number of interactions within the Heisenberg model. Decreasing the number of interaction parameters by creating a collective model for these systems significantly reduces the computational resources required for both first-principles calculations and spin dynamics simulations, allowing for a comparison with other systems. 

Spin spirals, such as the one depicted in Fig.~\ref{fig:CGT-lattices}(a), cover key magnetic states, like the ferromagnetic (FM), the antiferromagnetic (AFM), and the Néel state (see Fig.~\ref{fig:CGT-lattices}(g)), and can be computed in DFT within 
the chemical unit cell \cite{Kurz2004}. They describe a spatial rotation of magnetic moments $\mathbf{m}_i=(\cos(\mathbf{q}\cdot \mathbf{r}_i),\sin(\mathbf{q}\cdot \mathbf{r}_i),0)$ by a spin spiral vector $\mathbf{q}$. Spin spirals form the complete set of eigenmodes of the 
classical Heisenberg model on a periodic lattice. If there is only one magnetic atom inside the chemical unit cell, the spin spiral energy dispersion $E(\mathbf{q})$ repeats for each BZ, see Fig.~\ref{fig:CGT-lattices}(d). Only the irreducible wedge of the BZ needs to be sampled in order to access all spin spiral states.

In our collective atomistic model, the CGT and FGT layers are each modeled by a single species of magnetic atoms, which replicate the energy dispersions $E(\mathbf{q})$ in each layer by fitting exchange and DMI parameters to DFT 
total energy data. 

\subsubsection*{1. Intralayer interactions of FGT}

The three Fe layers of FGT (comprising a total of 9 Fe atoms, 3 Fe atoms from top view) are modeled by a hexagonal lattice, which contains three magnetic atoms in the chemical unit cell that cover the positions of the Fe atoms (see Fig.~\ref{fig:CGT-lattices}(b). Note that such a collective 2D model has been successfully used in our previous work \cite{Dongzhe2022_fgt, Dongzhe_PRB2024}, which predicts Curie temperatures in good agreement with the literature for FGT family materials. 
This lattice is three times denser than a hexagonal monolayer with just one atom per unit cell, and can be obtained by rotating the latter by 30$^\circ$ and scaling it with a factor of $\frac{1}{\sqrt{3}}$. The presence of more magnetic atoms in the chemical unit cell also has important implications for the space of possible spin states: while the lattice becomes three times smaller than the monolayer system, the space of possible spin spiral states becomes three times bigger, as can be seen in Fig.~\ref{fig:CGT-lattices}(e). In order to reach all new high symmetry states (Fig.~\ref{fig:CGT-lattices}(h)), longer paths of the $\mathbf{q}$ vector need to be taken into account. This model can transform complex multilayer interactions into simple monolayer interactions while preserving the same physics provided by $E(\mathbf{q})$.

\subsubsection*{2. Intralayer interactions of CGT}

The Cr atoms in the CGT layer form a honeycomb lattice with two atoms in the chemical unit cell. It can be obtained from the dense hexagonal lattice used for the FGT by removing one of the three sublattices, see Fig.~\ref{fig:CGT-lattices}(c). Spin spirals in this honeycomb model still use the large symmetry zone shown in Fig.~\ref{fig:CGT-lattices}(e), and the interaction parameters between the dense hexagonal lattice model and the honeycomb model can be related to each other.

In Fig.~\ref{fig:CGT-lattices}(f), the nearest neighbors in both models can be seen. For each of the three high-symmetry directions $\Delta\mathbf{R}_i$, an atom in the dense hexagonal layer model has two nearest neighbors, each of which contributes the energy $J_1^\text{hex}\cos(\Delta\mathbf{R}_i\cdot\mathbf{q})$. On the honeycomb lattice model, there is only one nearest neighbor for every $\Delta\mathbf{R}_i$, which also produces $J_1^\text{hon}\cos(\Delta\mathbf{R}_i\cdot\mathbf{q})$. Therefore, the energy dispersions of both models coincide for next neighbors if  $J_1^\text{hon}=2J_1^\text{hex}$. For higher orders of neighbors, there are shells with 50\% vacancies in the honeycomb lattice compared to the dense hexagonal, so $n_{\sigma}^{\text{hex}}=2n_{\sigma}^{\text{hon}}$, and some that have the same number of neighbors $n_{\sigma}^{\text{hex}}=n_{\sigma}^{\text{hon}}$. Therefore, the fitting of the CGT can be done with the dense hexagonal model, which has its parameters $C_{\sigma}^{\text{hex}}=J_{\sigma}^{\text{hex}}, D_{\sigma}^{\text{hex}}$, for exchange and DMI, weighted with the ratio $ n_{\sigma}^{\text{hex}}/n_{\sigma}^{\text{hon}}\in \{1,2\}$. This effectively comes down to the procedure
\begin{equation}\label{eq:hex_hon_constant_conversion}
    \arraycolsep=1.6pt\def\arraystretch{1.4}
    C_{\sigma}^{\text{hon}} = \left\{\begin{array}{cl}
        C_{\sigma}^{\text{hex}} &{\rm if }
        ~~ n_{\sigma}^{\text{hex}} = n_{\sigma}^{\text{hon}}\\
        2C_{\sigma}^{\text{hex}} &{\rm if }~~ n_{\sigma}^{\text{hex}} = 2n_{\sigma}^{\text{hon}}
    \end{array}\right. ~.
\end{equation}

\subsubsection*{3. Interlayer interactions between FGT and CGT}

The interlayer interactions ($E_{\perp}(\V{q})$) can be determined by calculating the energy difference between the total energy of the magnetic bilayer system and the sum of the intralayer interactions of FGT and CGT as follows 
\begin{equation}\label{eq:interlayer}
E_{\perp}(\V{q}) = E_{\text{FGT/CGT}}(\V{q}) - E_{\text{FGT}}(\V{q})-E_{\text{CGT}}(\V{q})
\end{equation}

By fitting $E_{\perp}(\V{q})$ with interlayer basis functions, we obtain $J_{\text{inter}}^{\perp, \sigma}$ and $D_{\text{inter}}^{\perp, \sigma}$ where $\sigma$ denotes the shell index. Since the two layers are coupled via van der Waals interactions, the coupling between FGT and CGT is relatively weak, as confirmed in our previous work \cite{Moritz_bimeron2025}. Therefore, to simplify the problem, we ignored the interlayer contribution in our atomistic spin simulations, as will be shown in Sec. \ref{results}.

The magnetic interactions obtained for different lattices are listed in Table~\ref{table_dmi}. These parameters are employed in our atomistic spin simulations to investigate soliton properties, such as their size and energy barriers.

\subsection*{D. Geodesic nudged elastic band method}
In this work, we investigate the stability of metastable soliton states by means of their energy barrier, which prevents them from collapsing into the field-polarized state, such as the FM state. This energy barrier $\Delta E$, also referred to as the activation energy of the annihilation event, is computed as the energy difference between the saddle point (SP) of the transition and the initial metastable soliton.
\begin{equation}
\Delta E = E_\text{SP} - E_\text{soliton}~.
\end{equation}
Local minimum energy configurations are obtained with the velocity projection optimization (VPO) algorithm \cite{bessarab2015method}, which is also employed as optimization scheme within the GNEB method used for the computation of minimum energy paths (MEP) between two of those minima based on the Hamiltonian in Eq.~(\ref{spin_model}).
For the scope of this work, the initial state always resembles the soliton, and the final state resembles the FM state. From the MEP, the SP, defined as the configuration of highest energy along the MEP, is obtained by the climbing image method \cite{henkelman2000}. These simulations were performed using $120 \times 120$ atomic sites, a convergence torque of $10^{-8}$ eV, and 50 images for the MEP.

\begin{figure}[t]
    \centering
    \includegraphics[width=0.95\linewidth]{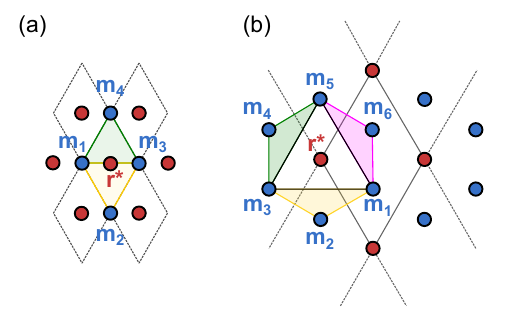}
    \caption{(a) Triangulation of the hexagonal lattice and (b) the honeycomb lattice for the definition of topological density $q$. The positions of the magnetic moments are depicted in blue, while the dual lattice points are shown in red. 
}
    \label{fig:topo_charge}
\end{figure}

\subsection*{E. Calculation of the topological charge density}

In the investigated material, all solitons can be described by a magnetization function $\mathbf{m}(\vec{r}):~\mathbb{R}^2\to\mathbb{S}^2$. For quantifying their topology, we consequently compute the degree of mapping $Q\in\mathbb{Z}$ for the corresponding homotopical group $\pi_2(\mathbb{S}^2)$
\begin{equation}\label{eq:topological_charge}
    Q=\frac{1}{4\pi}\int_{\mathbb{R}^2}\mathbf{m}\cdot\left(\frac{\partial \mathbf{m}}{\partial x}\times\frac{\partial \mathbf{m}}{\partial y}\right)~\mathrm{d}^2\mathbf{r}~,
\end{equation}
which, in the following, we will refer to as the topological charge. For a discrete lattice model, this charge has the form~\cite{berg1981}:
\begin{equation}\label{eq:topological_charge_dens}
    Q=\sum\limits_{\mathbf{r^*}}q(\mathbf{r}^*),
\end{equation}
where the topological density $q(\mathbf{r}^*)$ is defined at the vertices $\mathbf{r}^*$ of the dual lattice, the hexagonal (FGT) and honeycomb (CGT) lattices, respectively. 
For the hexagonal lattice (see Fig. \ref{fig:topo_charge}(a)) of the FGT layer, this density is a function of the four magnetic moments $\mathbf{m}_1$, $\mathbf{m}_2$, $\mathbf{m}_3$, and $\mathbf{m}_4$ at the lattice sites closest to $\mathbf{r}^*$:
\begin{align}\label{eq:charge_density_hex}
    q(\mathbf{r}^*)=\frac{1}{4\pi}\left(\mathcal{A}(\mathbf{m}_1,\mathbf{m}_2,\mathbf{m}_3)+\mathcal{A}(\mathbf{m}_1,\mathbf{m}_3,\mathbf{m}_4)\right).
\end{align}
Here, $\mathcal{A}(\mathbf{m}_1,\mathbf{m}_2,\mathbf{m}_3)$ denotes the signed area of the spherical triangle with corners $\mathbf{m}_1,\mathbf{m}_2,\mathbf{m}_3$.

For the honeycomb lattice of the CGT layer, each dual lattice point $\mathbf{r}^*$ is surrounded by six magnetic moments placed at the vertices of a hexagon. To define the spherical triangles, we triangulate the lattice, yielding one large main triangle and three smaller triangles. For each of these triangles, the topological density can be computed and assigned to the dual lattice points of the triangulated lattice. The topological density assigned to the center $\mathbf{r}^*$ of the honeycomb (see Fig. \ref{fig:topo_charge}(b)) is then:
\begin{align}\label{eq:charge_density_hon}
\begin{split}
    q(\mathbf{r}^*)=\frac{1}{4\pi}(&\mathcal{A}(\mathbf{m}_1,\mathbf{m}_5,\mathbf{m}_3)+\mathcal{A}(\mathbf{m}_1,\mathbf{m}_3,\mathbf{m}_2)\\
    &+\mathcal{A}(\mathbf{m}_3,\mathbf{m}_5,\mathbf{m}_4)+\mathcal{A}(\mathbf{m}_5,\mathbf{m}_1,\mathbf{m}_6)).
\end{split}
\end{align}

\section{RESULTS AND DISCUSSIONS}
\label{results}

In this section, we begin by discussing the validity of the fitting method, then present the first-principles results, and finally show the atomistic spin simulation results for FGT and CGT, respectively.

\begin{figure}[b]
    \centering
    \includegraphics[width=0.95\linewidth]{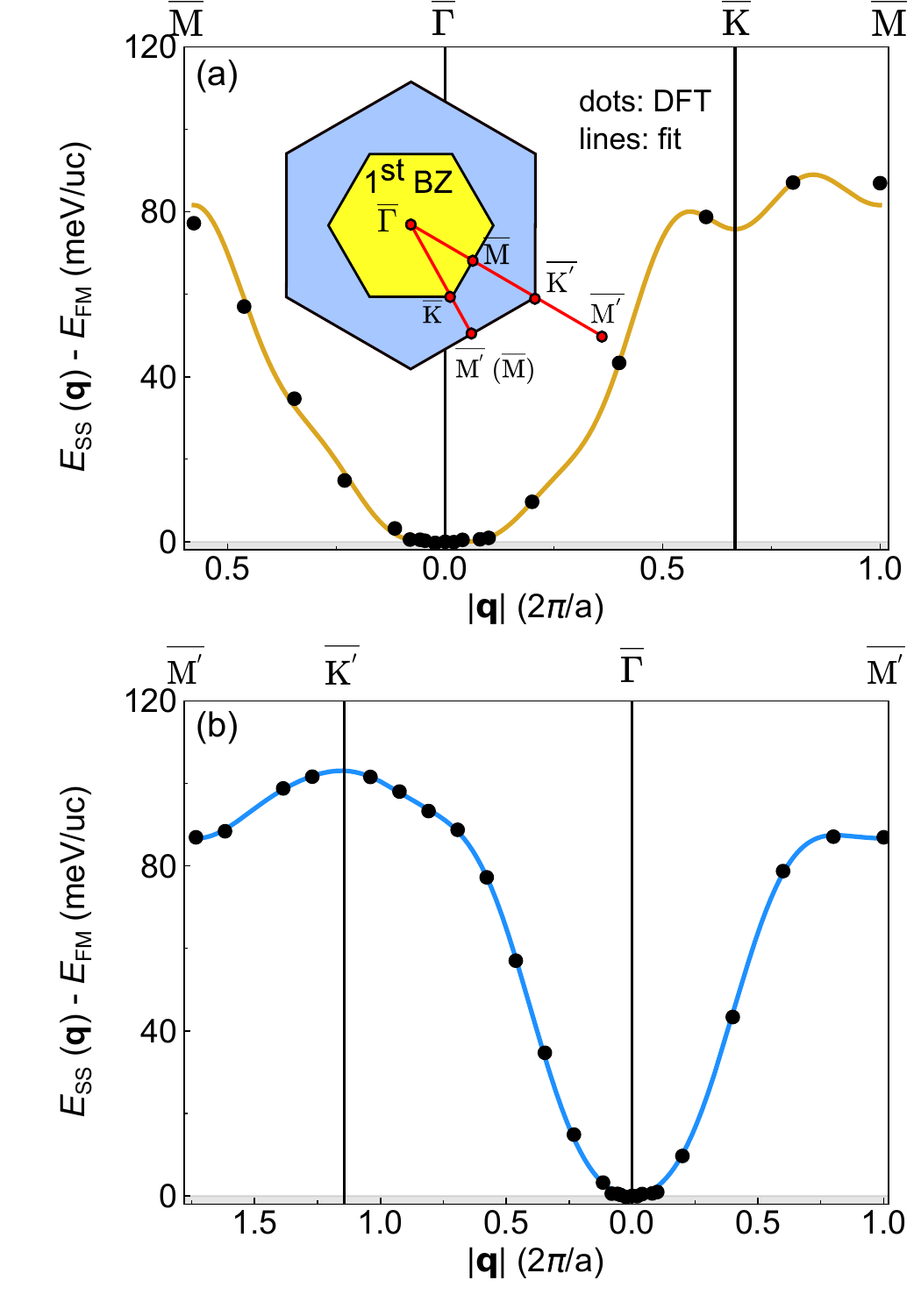}
    \caption{Comparison between two fitting schemes for the exchange constant of CGT. (a) DFT total energies (dots), with $\V{q}$ restricted to the first BZ, are fitted using the hexagonal monolayer model (cf.~Fig.~\ref{fig:CGT-lattices}(a) and (d)). As an inset, the extended BZ is shown with high-symmetry directions, $\overline{\Gamma \text{M}}$, $\overline{\Gamma \text{KM}}$, $\overline{\Gamma \text{M}^{'}}$, and $\overline{\Gamma \text{K}^{'}\text{M}^{'}}$. The first BZ is marked in yellow. (b) Same as in (a), but with DFT data sampled in the extended BZ and fitted using the CGT model (cf.~Fig.~\ref{fig:CGT-lattices}(c) and (f)).
}
    \label{fig:CGT-latticesfit}
\end{figure}

\begin{figure*}[t]
    \centering
    \includegraphics[width=1.0\linewidth]{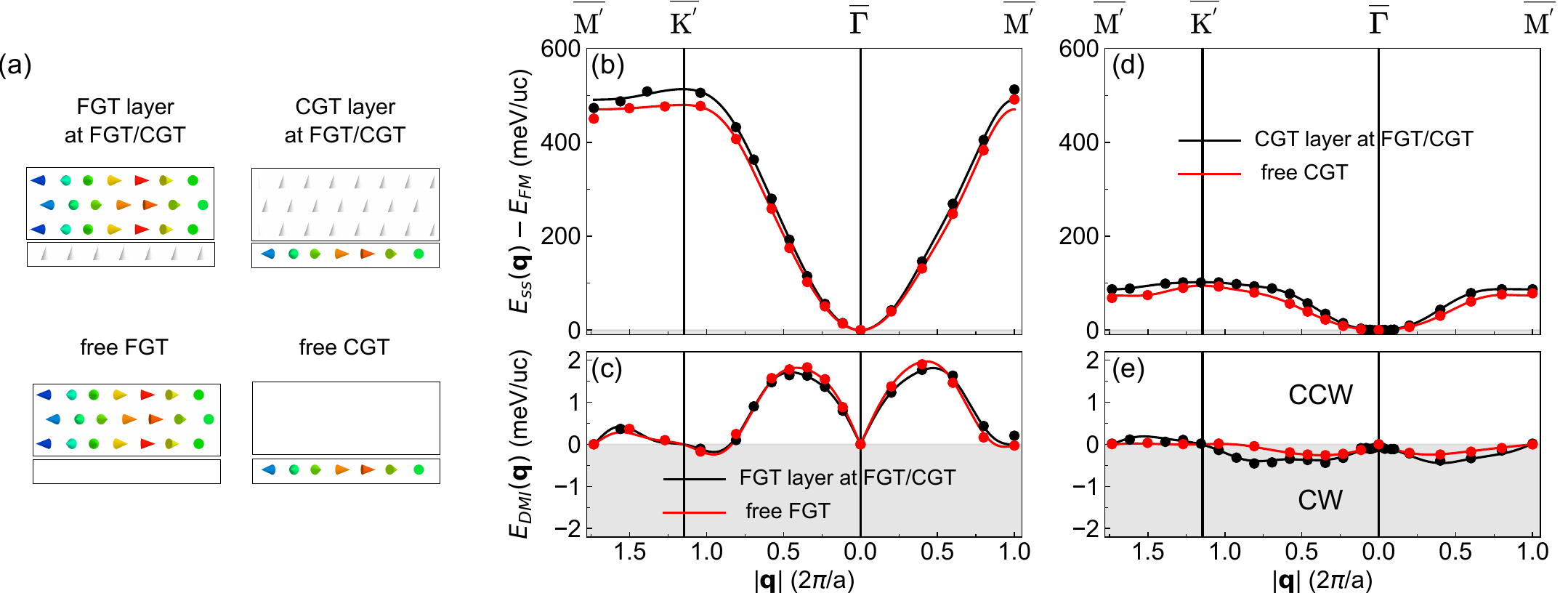}
    \caption{
Energy dispersion of spin spirals in the extended $\V{q}$ zone for the FGT and CGT layers at FGT/CGT. 
(a) Four spin spiral configurations used to calculate the energy dispersions. 
The top three layers represent the Fe atoms in the FGT layer, while the bottom layer represents the Cr atoms in the CGT layer.
(b-e) Spin spiral energy dispersions without SOC, $E_{\text{SS}}(\V{q})$ (upper panels), and SOC-induced DMI energy, $E_{\text{DMI}}(\V{q})$ (lower panels), for the FGT layer ((b) and (c)), and the CGT layer ((d) and (e)). Black circles correspond to the DFT total energies for FGT or CGT layers at FGT/CGT, while red circles are the corresponding DFT data for free FGT or CGT layers but with structural deformations as in FGT/CGT. All energies are given respect to the FM state at $\V{q} = 0$. Note that positive and negative DMI energy contributions represent clockwise (CW) and counter-clockwise (CCW) cycloidal spin spiral configurations. As we have verified, the ground state of both the FGT and the CGT layer is the FM state.}
    \label{spin-spiral}
\end{figure*}

\subsection*{A. Validation of the Heisenberg model fit}

To validate the new fitting scheme presented in Sec. \ref{fitt_largeQ}.C, we performed a test on the honeycomb lattice of CGT.
We plot in Fig.~\ref{fig:CGT-latticesfit} the fitting of the spin spiral dispersion with models based on a simple monolayer lattice (Fig. \ref{fig:CGT-latticesfit}(a)) and the dense hexagonal lattice (Fig. \ref{fig:CGT-latticesfit}(b)) used in the rest of this work on the example of the CGT layer. For the monolayer model, only spin spirals in the first BZ lead to different energy contributions. At the $\overline{\mathrm{M}}$ point, this results in a visible inconsistency in the energy dispersion: the two $\overline{\text{M}}$ points along $\overline{\Gamma \text{M}}$ and $\overline{\Gamma \text{KM}}$ are not degenerate (yellow hexagon in the inset of Fig.~\ref{fig:CGT-latticesfit}(a)). This is because the monolayer model does not consider the influence of interactions of atoms at a scale smaller than the chemical unit cell. As a result, the fitting shows a significant deviation, particularly along $\overline{\text{KM}}$ (see Fig.~\ref{fig:CGT-latticesfit}(a)). In contrast, the dense hexagonal monolayer accounts for these interactions and permits larger spin-spiral vectors $\mathbf{q}$, leading to different results. The two $\overline{\mathrm{M'}}$ points in the extended $\mathbf{q}$ zone (blue hexagon in the inset of Fig.~\ref{fig:CGT-latticesfit}(a)) are degenerate, as expected. The fit line shows excellent agreement with the DFT data.

\subsection*{B. Spin spiral dispersions}

To obtain the pair-wise Heisenberg exchange and DMI constants required for the atomistic spin model, we have performed spin spiral calculations \cite{Kurz2004,Heide2009,dupe2014tailoring,Zimmermann2014} via DFT. This approach enables us to explore a broad region of the magnetic phase space. The energy dispersion neglecting SOC, $E_{\text{ss}}(\V{q})$, is calculated via the \textsc{FLEUR} code
based on the generalized Bloch theorem \cite{Kurz2004}. 
The Heisenberg exchange constants are obtained by fitting the
self-consistently converged homogeneous flat spin spiral dispersions without SOC, $E_{\text{ss}}(\V{q})$. 
In the mapping to the Heisenberg model (cf.~Eq.~(\ref{spin_model})), we assume that the spin moments are constant with respect to $\V{q}$. This approximation is well justified for the CGT layer for all spin spiral vectors $\V{q}$ and for the FGT layer in the vicinity of the FM state as seen from the DFT calculations (see Appendix B).
As we have verified, the interlayer interactions between FGT and CGT are rather weak \cite{Moritz_bimeron2025}. Therefore, in the following, we focus only on the intralayer interactions within each layer. However, it is worth noting that our model includes the hybridization effect at the interface between FGT and CGT (see Appendix C).

Fig.~\ref{spin-spiral} displays the energy dispersion $E(\V{q})$ of flat homogeneous spin spirals per unit cell obtained via DFT
using the generalized Bloch theorem \cite{Kurz2004}. Here, $E(\V{q})$ is calculated along high-symmetry directions $\overline{\Gamma \text{M}^{'}}$ and $\overline{\Gamma \text{K}^{'}\text{M}^{'}}$ of the extended 2D hexagonal BZ shown in Fig.~\ref{fig:CGT-lattices}(e). The $\overline{\Gamma}$ point corresponds to the FM state, the $\overline{ \text{M}^{'}}$ point to the row-wise AFM state, and the $\overline{ \text{K}^{'}}$ point to the N\'eel-state with 120$^{\circ}$ angle between adjacent spins (Fig.~\ref{fig:CGT-lattices}(g)).

We first focus on intralayer magnetic interactions in the FGT layer. To extract only the Fe-Fe interaction parameters from DFT, we consider spin spirals in which only the Fe spins rotate, while the Cr spins in the CGT layer are in the FM state and perpendicular to the Fe spins 
(see upper eft panel in Fig.~\ref{spin-spiral}(a)). Without SOC, the ground state is the FM state ($\overline{\Gamma}$ point). Upon including SOC \cite{Heide2009}, DMI arises due to broken inversion symmetry at FGT/CGT. The DMI contribution to the dispersion, $E_{\text{DMI}}(\mathbf{q})$, is computed within first-order perturbation theory starting from the self-consistent spin spiral state \cite{Heide2009}. The FGT layer favors a CCW rotational sense, as seen from the calculated positive energy contribution, $E_{\text{DMI}}(\V{q})$, due to SOC to the dispersion of cycloidal spin spirals (Fig.~\ref{spin-spiral}(c)). 

Next, we investigate the intralayer interactions in the CGT layer (i.e., we fix all Fe spins to the FM state and rotate only the Cr spins, cf.~upper right panel in Fig.~\ref{spin-spiral}(a)). Again, we find the FM state to be energetically lowest (Fig.~\ref{spin-spiral}(d)). 
However, a much smaller nearest-neighbor Heisenberg exchange is observed compared to the one at the FGT layer, as clearly seen from the much-reduced energy difference between the 
$\overline{\Gamma}$ and the $\overline{\text{M}^{'}}$ point
(for exchange constants, see Appendix A). Upon including DMI, the CGT layer favors cycloidal spin spirals with a clockwise (CW) rotational sense, i.e., $E_{\text{DMI}}(\V{q})<0$ (Fig.~\ref{spin-spiral}(e)), which is opposite to that of the FGT layer. 

\begin{table}[t]
        \caption{\label{DMI_micro} Micromagnetic DMI amplitudes $|D|$ for the FGT and CGT layers at FGT/CGT are compared with those of other FGT vdW heterostructures reported in the literature.}
	\centering
	\scalebox{0.95}{
		\begin{tabular}{cccccccccccccccccc}
			\hline\hline
			\multicolumn{1}{c}{ } & \multicolumn{1}{c} {~~~$|D|$ (mJ/m$^2$)~~~}  \\ 
   \hline
            ~~~pristine FGT~~~  &  0.00 \\
            ~~~strained FGT/Ge (theory) \cite{Dongzhe2022_fgt}~~~ & 1.50 \\
             ~~~FGT/In$_2$Se$_3$ (theory) \cite{Huang2022}~~~&  0.28 \\
            ~~~FGT-Li (theory) \cite{fgt-Li}~~~&  1.41 \\
            ~~~FGT/WTe$_2$ (experiment) \cite{wu2020neel}~~~&  1.00 \\
            ~~~FGT layer at FGT/CGT (experiment) \cite{wu2022van}~~~&  0.18$\sim$0.44 \\
            ~~~FGT layer at FGT/CGT (this work, theory) ~~~& 0.13 \\
            ~~~CGT layer at FGT/CGT (this work, theory) ~~~& 0.10 \\
			\hline
   \hline
			\end{tabular}}
		\end{table}		

We also estimate the corresponding micromagnetic DMI coefficient $D$, which is often used to interpret experimental data. The coefficient $D$ is given by:
\begin{equation}
D = \frac{3\sqrt{2}d_{\parallel}}{N_{\text{FM}} a^2},
\end{equation}
where \(a\), \(N_{\text{FM}}\), and $d_\parallel$ are the lattice constant, the number of ferromagnetic layers, and the in-plane component of the DMI constant in the nearest-neighbor (NN) approximation, respectively. We used $a = 0.693$ nm~for both FGT and CGT layers, and we took $N_{\text{FM}} = 3$ and $N_{\text{FM}} = 1$ for the $D$ calculations of the FGT and CGT layers, respectively.

\begin{figure*}[!tbp]
 	\centering
 	\includegraphics[width=1.0\linewidth]{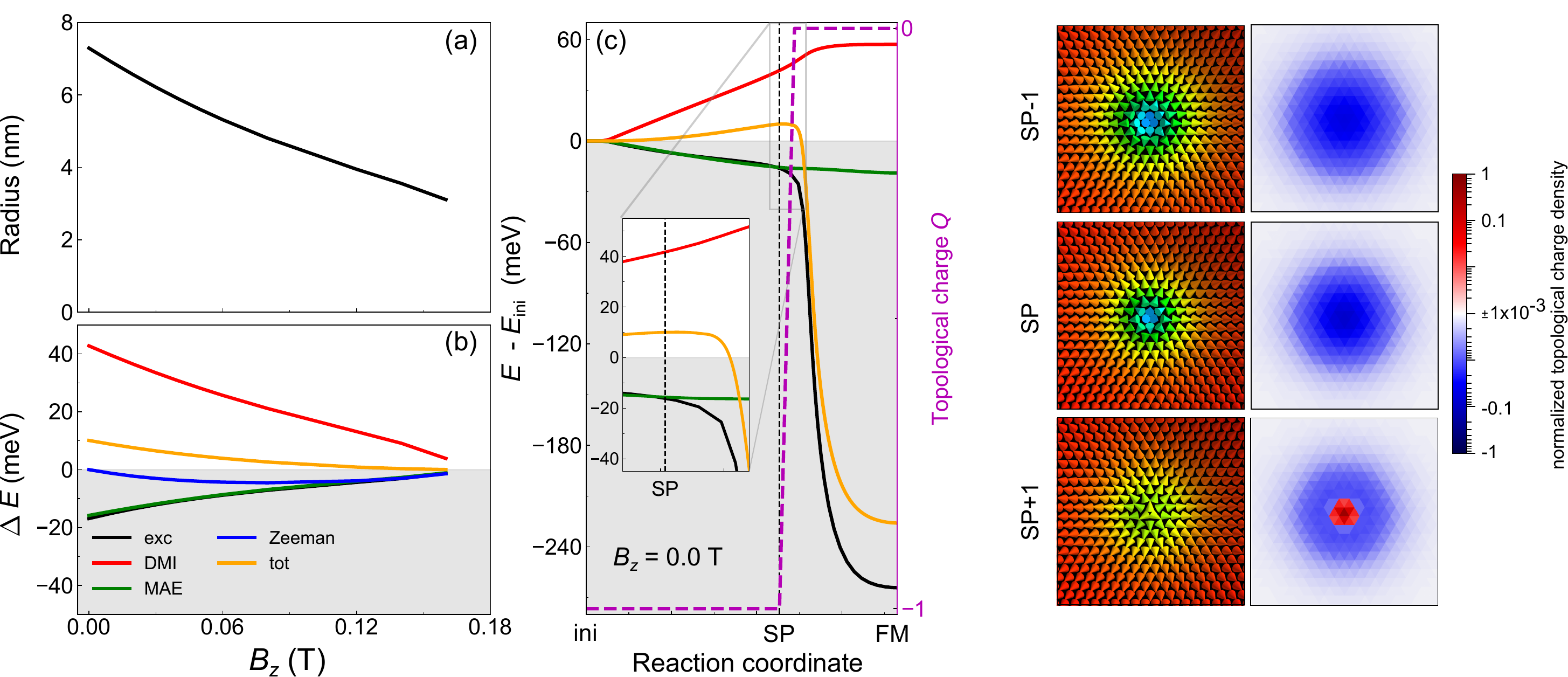}
 	\caption{\label{GNEB_skyrmion} Radius, energy barrier, and collapse mechanism of magnetic skyrmions in the FGT layer. (a) Skyrmion radius and (b) total energy barrier $\Delta E$ for skyrmion collapse decomposed into the contributions from different interactions (see legend)   
    as a function of the applied out-of-plane magnetic field ($B_z$). (c) Minimum energy path (MEP) for the collapse of the skyrmion (initial) state into the FM (final) state through the saddle point (SP) at $B_z$ = 0.0 T. The topological charge (open circles,  right axis) is plotted versus the reaction coordinate. The corresponding spin structures and topological charge densities from Eq.~(\ref{eq:charge_density_hex}) before (SP-1), after (SP+1), and at the saddle point (SP) are shown on the right side of panel (c). For enhanced visibility, the topological charge is normalized on $[-1, 1]$ and its absolute value in the interval $q\in[-1,-1 \times 10^{-3})\cup(1 \times 10^{-3}, 1]$ is displayed logarithmically, including the respective sign. The sign switching of the topological charge density in the center, between images SP and SP+1, marks the disappearance of the total topological charge. Note that the skyrmion collapse occurs via radially symmetrical shrinking, until a Bloch-point-like configuration of the center spins is reached.
 	}
\end{figure*}

The calculated micromagnetic DMI amplitude for the FGT layer is found to be about 0.13 mJ/m$^2$, which is in reasonable agreement with the experimental value of 0.18–0.44 mJ/m$^2$ \cite{wu2022van}. The $|D|$ for the CGT layer is about 0.10 mJ/m$^2$. The comparison with other DMI amplitudes in various FGT heterostructures, in which magnetic skyrmions are observed, is summarized in Table \ref{DMI_micro}. It is clear that the DMI in both FGT and CGT layers is relatively small, indicating that the formation of topological solitons does not purely rely on DMI.

As demonstrated in our previous work \cite{Dongzhe2022_fgt}, the DMI is sensitive to the degree of inversion-symmetry breaking. To quantify the hybridization effect at the interface in the magnetic interactions and to distinguish it from the structural relaxation effect, we calculated the exchange and DMI constants for free-standing FGT and CGT layers, incorporating the deformation induced by the FGT/CGT heterostructure (see red curves in Fig. \ref{spin-spiral}). We find that the structural effect strongly dominates in both FGT and CGT layers. The hybridization effect is comparatively small, though slightly more pronounced in the CGT layer. This is also reflected in the small interlayer exchange constants between the FGT and CGT layers \cite{Moritz_bimeron2025}.

In addition, we find that the easy axis of the FGT layer is out-of-plane, $K = 9.81$~meV/uc, while it is in-plane for the CGT layer, $K = -0.26$~meV/uc. Together with the weak interlayer coupling, this gives rise to the possibility of engineering multiple topological spin textures in one system.

\subsection*{C. Skyrmion stability and collapse mechanism in FGT}

Based on the spin Hamiltonian, Eq.~(\ref{spin_model}), parameterized from DFT, we have performed atomistic spin simulations to explore the possibility of inducing and stabilizing chiral spin textures at FGT/CGT. Since the hybridization between FGT and CGT is rather small, we treat them in our spin model as separate layers and neglect the interlayer coupling.

We first discuss the spin structures obtained in the FGT layer. We do not find topologically non-trivial states emerging in our simulations with the magnetic parameters obtained by DFT. This is due to a very large MAE ($K$ = 1.09 meV/Fe) predicted by DFT. However, the MAE of FGT can be significantly reduced by doping or temperature, as demonstrated in many experiments \cite{tan2018hard,wang2020modifications,Park2020_nl}. Therefore, we performed atomistic spin simulations with a reduced value of the MAE of $K/100$, as also proposed in a recent theory work by Huang \textit{et al.}~\cite{Huang2022} for FGT heterostuctures. Our simulations predict the emergence of metastable isolated skyrmions (Néel-type) in the FM background. 

\begin{figure*}[htp]
    \centering
    \includegraphics[width=1.0\linewidth]{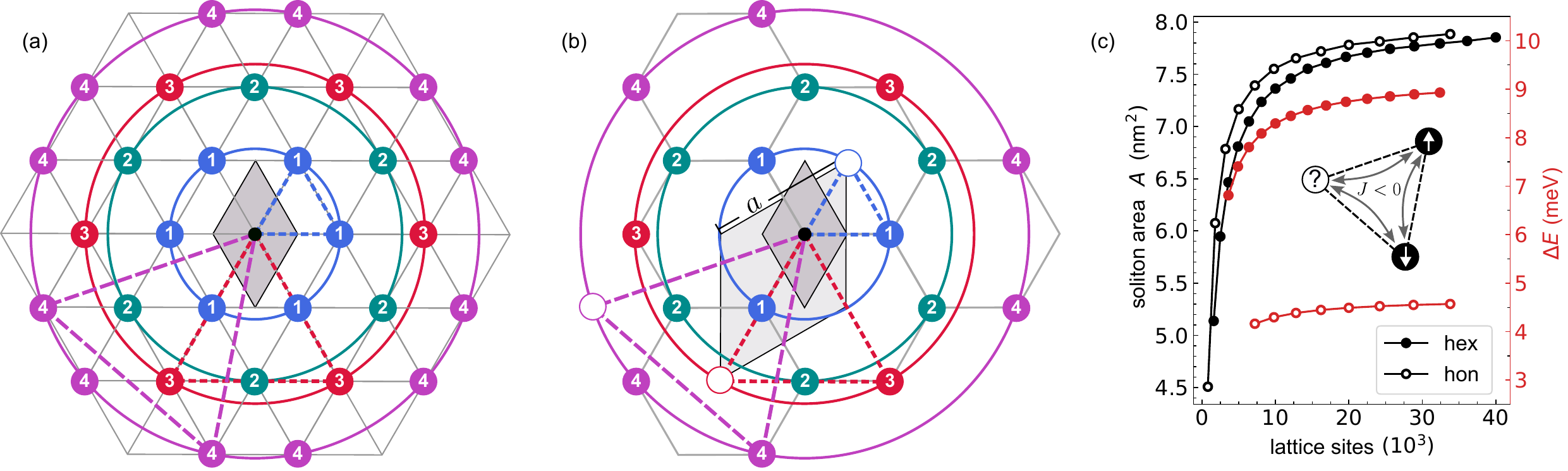}
    \caption{Shell-resolved interactions on (a) the hexagonal and (b) the honeycomb lattice. Here, the lattice constant $a=0.693~$nm for the real-world honeycomb lattice of CGT is drawn, which is larger by a factor of $\sqrt{3}$ compared to the lattice constant of the hexagonal lattice. It shall be illustrated how removing sites from the hexagonal lattice gives rise to the honeycomb geometry, by which only certain shells are affected. While the transformation of interactions in Eq.~(\ref{eq:hex_hon_constant_conversion}) conserves the energy dispersion and spin stiffness under this operation, the comparison of triangles drawn in (a) and (b) illustrates that the geometric frustration for the affected shells is lifted. The respective lattice sites, which, due to their absence, cause this lifting, are marked by empty circles. (c) Comparison of bimeron area $A$ (black dots, left $y$-axis) and the energy barrier (red dots, right $y$-axis) as functions of system size for the two lattice geometries.
    Note that for both lattice geometries, the soliton area is estimated with respect to the lattice constant $a$ of the honeycomb lattice, keeping the scale ratios of (a) and (b) as drawn in this figure. Notably, the area on both geometries stays the same, but the energy barriers differ by a factor of $\sim2$.
    }
    \label{fig:hex_vs_hon}
\end{figure*}

For the characterization of the size of radial symmetric skyrmions in the FGT layer, we rely on their radius rather than their area.  Here, we deploy the usual vortex radius estimation \cite{bocdanov1994properties} based on the inflection point $\rho_c$ of the $\Theta$-profile, where $\ddot{\Theta}(\rho_c)=0$ and
\begin{equation}\label{eq:skyrmion_radius}
    R_{\text{sk}}= \rho_c - \frac{\Theta(\rho_c)}{\dot{\Theta}(\rho_c)}~.
\end{equation}

Remarkably, nanoscale skyrmions with radii ranging from approximately 3 to 8 nm are observed at small magnetic fields below 0.16 T down to zero field. As expected, their radii decrease rapidly with increasing magnetic field (Fig.~\ref{GNEB_skyrmion}(a)). Notably, these nanoscale skyrmions are much smaller than those reported in strained FGT/Ge \cite{Dongzhe2022_fgt} for the same value of MAE.

To quantify the stability of topological spin structures, we calculated the energy barriers for the collapse of isolated skyrmions in the FM background using the GNEB method~\cite{bessarab2015method,Malottki2019}. We find that skyrmions in the FGT layer exhibit an energy barrier of about 10~meV at zero field [Fig.~\ref{GNEB_skyrmion}(b)], which rapidly decreases with increasing magnetic field. We find that the metastable state of isolated skyrmions vanishes at $B_z \approx 0.16$~T. Our results indicate that the DMI is mainly responsible for skyrmion stability, whereas the exchange interaction and the MAE favor the FM state as shown in Fig.~\ref{GNEB_skyrmion}(b-c). This also explains why reducing the MAE is necessary to stabilize skyrmions. A similar scenario has been reported for strained FGT/Ge~\cite{Dongzhe2022_fgt}. In Fig.~\ref{GNEB_skyrmion}(c), we plot the MEP for the collapse of a single skyrmion into the FM state at $B_z = 0$~T. In the collection of images along the MEP, the Bloch point occurs between images SP and SP+1, where SP denotes the point of maximum energy with respect to the initial skyrmion state. Analysis of the spin structure further reveals the radial collapse mechanism for skyrmions~\cite{malottki2017,Malottki2019,muckel2021experimental}.

Here, we note that the skyrmion properties obtained for the FGT layer, such as the size and $\Delta E$, cannot be regarded as quantitative results, as we have adjusted only the MAE value in our atomistic spin simulations.

\subsection*{D. Magnetic solitons in CGT on hexagonal and honeycomb geometries}

In contrast to FGT, the CGT layer obeys an easy-plane anisotropy and therefore hosts vortex-bound states, e.g., bimerons and antibimerons at zero field, instead of skyrmions. A detailed analysis of the properties and stability of magnetic solitons, as well as their transformation under an external magnetic field in this material, can be found in our recent work \cite{Moritz_bimeron2025}. Instead, in the spirit of the comparison between hexagonal and honeycomb lattice presented in the previous sections, here we focus on soliton features that arise due to the differences between these two lattice models.

Both lattice geometries, including the shells of isotropic pairwise interactions, are illustrated in Fig.~\ref{fig:hex_vs_hon}(a-b). In the comparison between the two lattices, it becomes clear that the honeycomb lattice can be constructed by removing certain lattice sites from the hexagonal lattice. This is the basis for the conversion of interaction parameters introduced in Eq.~(\ref{eq:hex_hon_constant_conversion}) that leaves the energy of spin spirals invariant under removing lattice sites from the hexagonal lattice in order to obtain a honeycomb lattice. For spin spiral states $\mathbf{m}_{\mathbf{q}}$, characterized by a wave vector $\mathbf{q}\in\mathbb{R}^3$ and orientation vector $\mathbf{c}=\mathbf{R} +\mathrm{i}\mathbf{I}\in\mathbb{R}^3$, $\mathbf{R}\perp\mathbf{I}$ with $|\mathbf{R}|=|\mathbf{I}|=1/2$, the magnetization at each lattice site $\mathbf{r}_n$, $n\leq N$ takes the vector
\begin{align}\label{eq:spin_spiral_basis}
    \begin{split}
        \mathbf{m}_{\mathbf{q}}\ni\mathbf{m}_{\mathbf{q}}(\mathbf{r}_n) = \mathbf{c}\mathrm{e}^{\mathrm{i}\mathbf{q}\cdot\mathbf{r}_n} + \mathbf{c}^*\mathrm{e}^{-\mathrm{i}\mathbf{q}\cdot\mathbf{r}_n} 
    \end{split}
\end{align}
As utilized in the fitting procedure described in Sec. \ref{fitt_largeQ}.C, each spin spiral has an exchange and DMI energy, which is decoupled from other spin spirals. In our Hamiltonian, it reads
\begin{align}
    \begin{split}
        \frac{E(\mathbf{m}_{\mathbf{q}})}{N} 
        &= -\sum_\sigma J_\sigma \sum_{\mathbf{r}\in\sigma} 
        \cos(\mathbf{q} \cdot \mathbf{r}) \\
        &\ \quad - 2 \mathbf{I} \times \mathbf{R} 
        \sum_\sigma D_\sigma \sum_{\mathbf{r} \in \sigma} 
        (\mathbf{r} \times \hat{\mathbf{z}}) \sin(\mathbf{q} \cdot \mathbf{r})~.
    \end{split}
\end{align}
In addition to the decoupled energy, spin spirals also resemble a set of orthogonal magnetic states
\begin{equation}
    \mathbf{m}_{\mathbf{q}}\cdot \mathbf{m}_{\mathbf{q'}}= \sum_{n=1}^N\mathbf{m}_{\mathbf{q}}(\mathbf{r}_n)\cdot\mathbf{m}_{\mathbf{q}'}(\mathbf{r}_n) = 2N\delta_{\mathbf{q},\mathbf{q'}}
\end{equation}
which arises from Eq.~(\ref{eq:spin_spiral_basis}) in combination with the well-known generalized orthogonality relation $\sum_{n}\mathrm{e}^{\mathrm{i}(\mathbf{q}-\mathbf{q}')\cdot\mathbf{r}_n}=N\delta_{\mathbf{q},\mathbf{q}'}$. Thus spin spirals can be used to represent arbitrary spin states $\mathbf{m}$ in a basis that is linear in exchange and DMI
\begin{equation}
    E(\mathbf{m}) = E\left(\sum_{\mathbf{q}} \xi_{\mathbf{q}}\mathbf{m}_{\mathbf{q}} \right) = \sum_{\mathbf{q}} \xi_{\mathbf{q}}E(\mathbf{m}_{\mathbf{q}})
\end{equation}
with coefficients $\xi_{\mathbf{q}}\in\mathbb{R}$. Consequently, this means that the conversion between hexagonal and honeycomb lattices, which leaves the energy $E(\mathbf{m}_{\mathbf{q}})$ of spin spirals invariant, also preserves the energy $E(\mathbf{m})$ of arbitrary states, only if the composition $\{\xi_{\mathbf{q}}\}$ stays the same. However, since the composition can vary under conversion -- for example, removing all $-m_z$-spins from a hexagonal ferrimagnet, a 2Q state, leaves a $+m_z$-ferromagnet on a honeycomb, a 1Q state -- the properties of soliton configurations that are the superposition of many spin spirals, such as bimerons, are difficult to predict.

As pointed out in Ref.~\cite{Moritz_bimeron2025}, bimerons in CGT exhibit a non-radial symmetric shape. Thus, the ansatz for the radius estimation in Eq.~(\ref{eq:skyrmion_radius}) is not applicable. Instead, we determine the bimeron size in terms of the area $A$ that is enclosed by the angle ${\pi}/{2}$ of the magnetization against the background. Since it is known that solitons in easy-plane magnets are sensitive to system dimensions \cite{Moritz_bimeron2025}, we investigate their evolution with simulation box size. In this regard, due to the phenomenological connection between size and stability \cite{varentcova2020toward}, we also investigate the evolution of the energy barrier $\Delta E$, which is computed in the same way as for skyrmions in FGT. 

The results for bimerons at $B_z=0~$T are displayed in Fig.~\ref{fig:hex_vs_hon}(c). It can be seen that the area $A$ of bimerons on the honeycomb and hexagonal lattice (black dots) takes similar values for the same number of lattice sites in the simulation box. This is in good agreement with the conversion of interaction constants from Eq.~(\ref{eq:hex_hon_constant_conversion}), which is constructed to leave the spin spiral dispersion invariant and thus should also lead to a similar size of bimerons. However, this correspondence does not apply to energy barriers (red dots), as shown in Fig.~\ref{fig:hex_vs_hon}(c). Here, bimerons on the hexagonal lattice are protected by a barrier that is higher by a factor of $\sim2$ compared to bimerons on the honeycomb lattice.
Since the conversion of interaction constants we apply (cf. Eq.~(\ref{eq:hex_hon_constant_conversion})) leaves the micromagnetic spin stiffness $\mathcal{J}$ invariant when considering the actual number of lattice sites $n_{\sigma}$ in each shell $\sigma$, as well as their distance $|\mathbf{r}_{\sigma}|$
\begin{equation}
    \mathcal{J} = \frac{1}{2}\sum_{\sigma}n_{\sigma}^{\text{hex}}J_{\sigma}^{\text{hex}}|\mathbf{r}_{\sigma}|^2 = \frac{1}{2}\sum_{\sigma}n_{\sigma}^{\text{hon}}J_{\sigma}^{\text{hon}}|\mathbf{r}_{\sigma}|^2 ~,
\end{equation}
here, the difference in energy barriers has to originate from discrete lattice effects. 

\begin{figure*}[htp]
    \centering
    \includegraphics[width=1.0\linewidth]{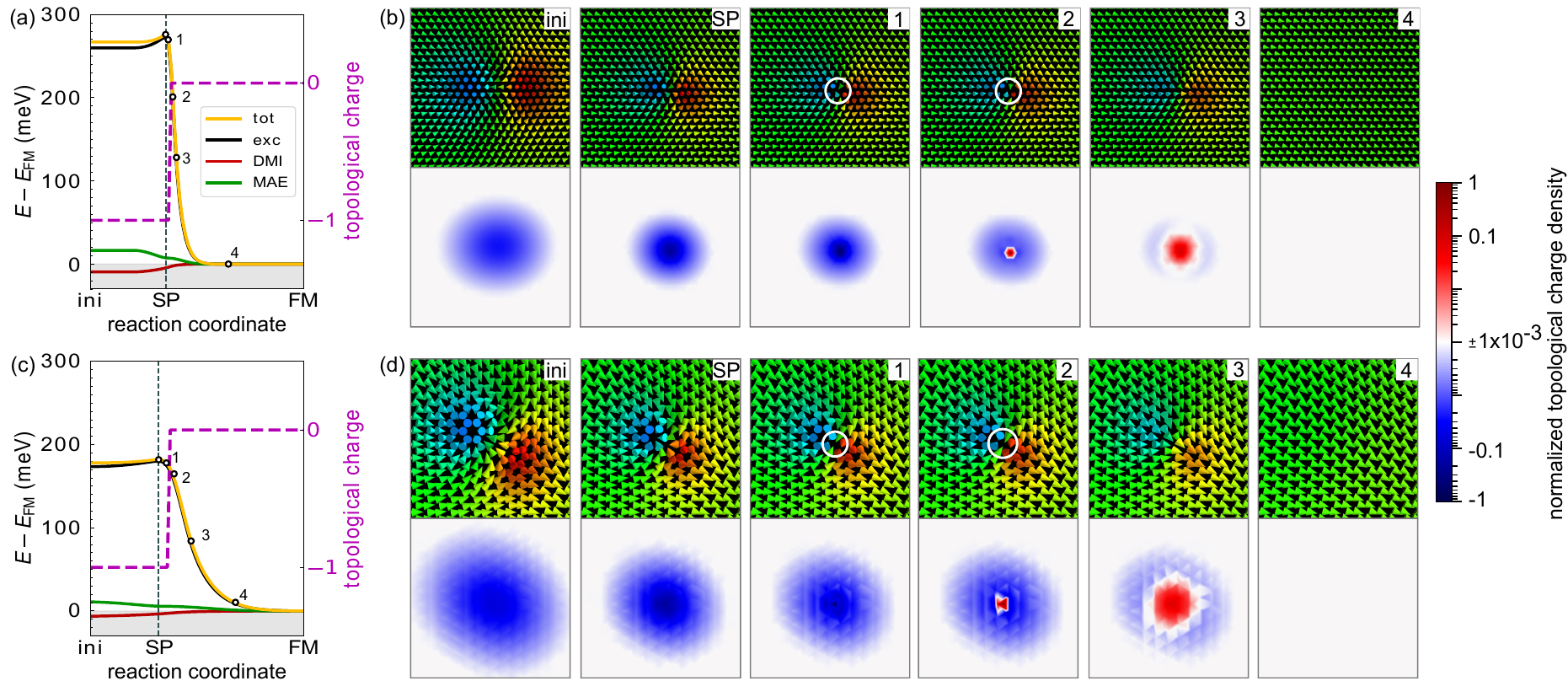}
    \caption{    
MEP and transition mechanism of bimeron collapse on hexagonal and honeycomb lattices at zero field. (a) Total energy (orange) of images along the MEP for the transition between the bimeron and the FM state on the hexagonal lattice. The energy is decomposed into exchange (black), DMI (red), and MAE (green). All energies are plotted with respect to the FM state. The topological charge (violet, dashed) of the images is plotted with respect to the right axis. (b) Selected spin textures (top panels) along the MEP. The numbering denotes the positions of the images as indicated in (a), and the white circles in images 1,2 indicate the position where the sign of the topological charge density is inverted. The corresponding topological charge densities are also shown (bottom panels). (c–d) Same as (a–b), but for the topological charge of the honeycomb lattice according to Eq.~(\ref{eq:charge_density_hon}). In both lattices, the SP does not coincide with the BP.
}
    \label{fig:gneb_hex_hon}
\end{figure*}

\begin{figure*}[!tbp]
	\centering
	\includegraphics[width=1.0\linewidth]{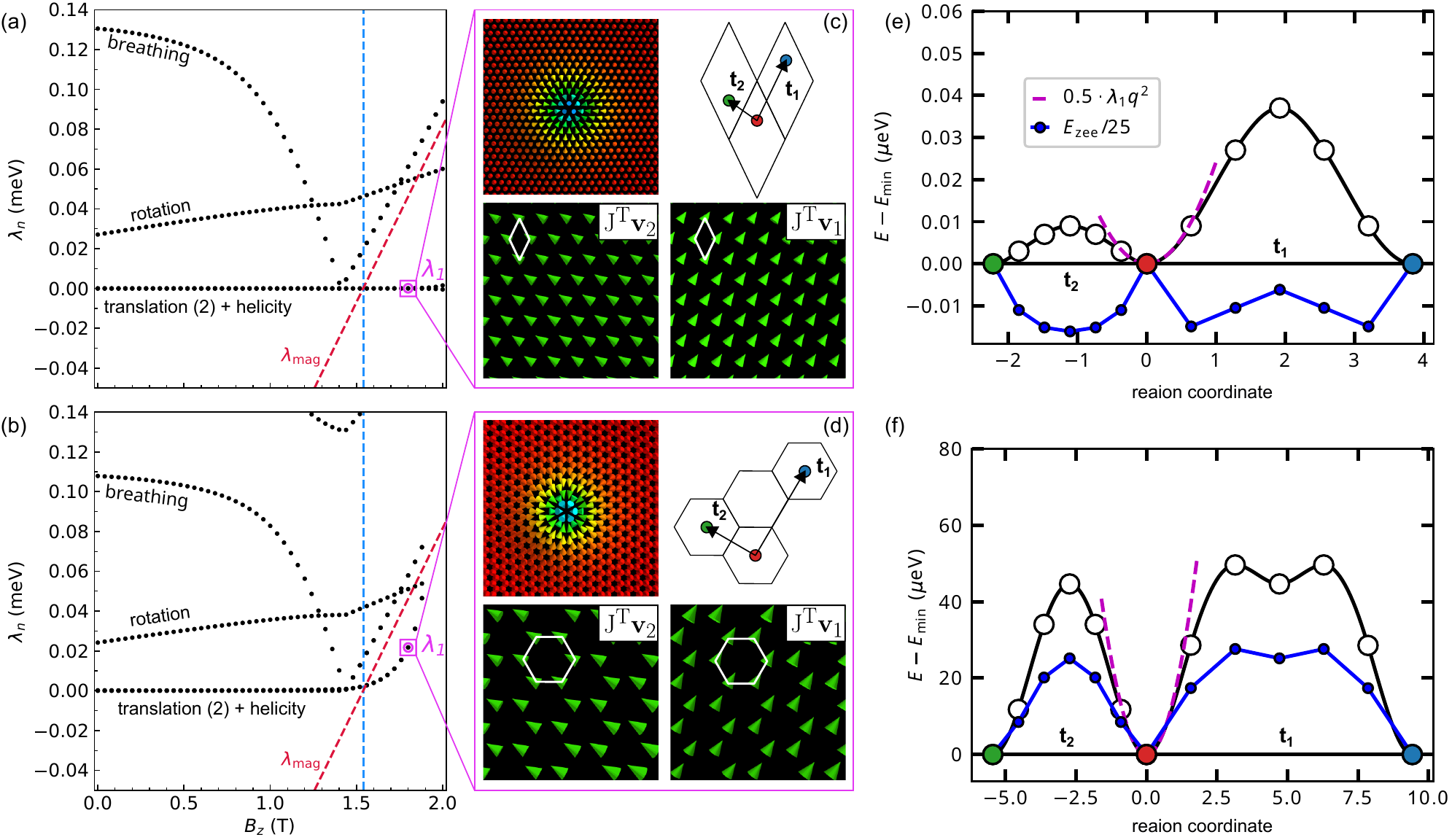}
	\caption{Lowest curvatures of the energy along Hessian eigenvectors for a skyrmion on a hexagonal lattice (a) and a honeycomb lattice (b). The magnon gap $\lambda_{\text{mag}}$ is indicated by a red line. The vertical blue line marks the position of $B_z^{\text{sat}}$, above which the background is polarized in the $ m_z$ direction. For the translation of the skyrmion, the Jacobian transformation of the corresponding eigenvector is shown in (c) for the hexagonal lattice and in (d) for the honeycomb lattice. MEP between translated skyrmions for $B_z=1.8$~T in the (e) hexagonal and (f) honeycomb lattices, respectively. The center of the skyrmion (red, blue, and green circles) is located at three positions shifted toward each other with respect to the directions $\vec{t}_1$ and $\vec{t}_2$ for the translation modes. The reaction coordinate is given in units of the geodesic distance between translated spin configurations. The harmonic approximation of the energy according to the lowest eigenvalue of the Hessian is depicted in magenta. The corresponding eigenvalues $\lambda_1$ are also marked in (a) and (b) for clarification of their origin. The contribution of the Zeeman energy is drawn in blue with a reduced scale (1/25) along the MEP.
    }
    \label{fig:translation}
\end{figure*} 

\subsubsection*{1. Geometric frustration}

The major difference that arises in the conversion from hexagonal to honeycomb lattice is the lifting of geometric frustration in certain shells. Here, geometric frustration describes that under an exchange interaction with constant $J<0$ there exists no energy-minimizing collinear alignment for spins on an isosceles triangle (cf. inset of Fig.~\ref{fig:hex_vs_hon}(c)). Instead, the competition leads to the formation of a non-collinear alignment. Exemplary triangles, on which geometric frustration is lifted by the loss of a lattice site under the conversion from hexagonal to honeycomb lattice, are drawn in Fig.~\ref{fig:hex_vs_hon}(a-b). 
The additional requirement of having a negative interaction constant $J<0$, in order for geometric frustration to be relevant (cf.~Table~\ref{table_dmi}), tells us that geometric frustration is lifted in shells with index 4 and 7. Besides that, there of course exist other non-isosceles triangles for triplets of spins that interact by two different exchange constants $J_i,J_j<0$, which also have differently frustrated contributions on both lattices.

Surprisingly, we find that the loss of frustration amounts to a decrease in the bimeron 
energy barrier by a factor $\sim 2$ on the honeycomb lattice, compared to the hexagonal geometry.
Note that a large effect of exchange frustration on the stability of isolated skyrmions in ultrathin transition-metal films has been previously reported \cite{malottki2017,meyer2019isolated}.

\subsubsection*{2. Bimeron collapse mechanism}

The energy decomposition of the MEP for the bimeron collapse at $B_z=0~$T into contributions from all interactions of the Hamiltonian in Eq.~(\ref{spin_model}) is shown in Fig.~\ref{fig:gneb_hex_hon}(a) for the hexagonal and in Fig.~\ref{fig:gneb_hex_hon}(c) for the honeycomb lattice. In both graphs, the energy is given with respect to the FM state, which is the final state of the transition. In this way, it becomes directly apparent that the energy difference between the skyrmion state and the FM is significantly larger on the hexagonal compared to the honeycomb lattice. The largest contribution to this difference stems from the exchange interaction due to exchange
frustration between short-range and long-range interactions (see Table \ref{table_dmi}), whereas the DMI and the MAE show minor deviations between the two lattice types. 

For the difference in energy barriers observed in Fig.~\ref{fig:hex_vs_hon}(c), the energy difference to the FM state is not important. The deviation should rather be attributed to differences in the spin constellation at the honeycomb and hexagonal saddle points, caused by the differences in geometric frustration. In Fig.~\ref{fig:gneb_hex_hon}(b-c), important images along the MEP are shown for both lattices, respectively. While the topological charge density, given below the respective spin textures, is very similar in both cases, the most evident difference is that the collapse mechanism in the vicinity of SP mostly involves the movement of two spins on the hexagonal lattice, but six spins on the honeycomb lattice (see in particular the spin textures inside the white circles). This collective behavior effectively lowers the cost of exchange energy and consequently the energy barrier. However, since the MEP is the result of a numerical optimization procedure, the exact relation between the loss of geometric frustration and variations in the collapse mechanism cannot be revealed. Rather, our results demonstrate that material-specific discretization impacts soliton stability, here by a factor of 2, and that the reliability of theoretical predictions thus stands and falls with the precision of models.

\subsubsection*{3. Skyrmion pinning energetics}

Another difference between solitons in 2D hexagonal and honeycomb geometries concerns their pinning to the lattice. In Fig.~\ref{fig:translation}(a) and (b), we compare the curvature along the low-energy deflections of skyrmions in CGT over the range of magnetic fields $B_z\in[0,2~\text{T}]$, which has also been investigated in Ref.~\cite{Moritz_bimeron2025}. The curvatures are obtained as the lowest eigenvalues $\lambda_n$ of the spherical constraint Hessian $\mathrm{H}\in\mathbb{R}^{2N\times2N}$ \cite{bessarab2012harmonic}, where $N$ is the number of lattice sites. By analyzing the respective eigenvectors $\mathbf{v}_n$ that solve 
\begin{equation}
    \mathrm{H}\mathbf{v}_n=\lambda_n\mathbf{v}_n
\end{equation}
we identify the respective degrees of freedom, which in the spectrum of interest are the breathing-, rotation-, helicity-, and translation modes. Away from exact values, we find qualitative agreement between these excitations on both lattices, with the exception of the translation mode. Note that beyond a magnetic field of 1.54~T, bimerons transform into skyrmion-like textures, as demonstrated in Ref.~\cite{Moritz_bimeron2025}. In the following, we will exemplarily focus on the skyrmion state at $B_z = 1.8$ T. We find that the curvature of the energy along the translation stays $\lambda\approx0$ on the hexagonal lattice, the curvature on the honeycomb lattice increases rapidly for $B_z>1.54~$T, beyond the field of critical coupling \cite{barton2020magnetic, Moritz_bimeron2025}. This means that moving the skyrmion comes along with increasing energy cost, so that the skyrmion gets pinned for sufficiently low temperatures.

For better visualization, we display the two translation eigenvectors projected onto the Jacobian $\mathrm{J}\in\mathbb{R}^{3N\times3}$ of the respective skyrmion state $\mathbf{m}\in\mathbb{R}^{3N}$. Note that since $\mathbf{v}_n$ is a vector in the 2$N$-dimensional tangent space of the configuration space of magnetic textures, the visualization requires a projection in the embedding $\mathbb{R}^3$. Consequently, using $\Tilde{\mathbf{v}}_n=\mathrm{P}\mathbf{v}_n\in\mathbb{R}^3$, this operation gives us the deflection of the magnetization in the direction of the eigenvector, because
\begin{equation}
    (\Tilde{\mathbf{v}}_n\cdot\nabla)\mathbf{m} = \mathrm{J}^{\text{T}}\Tilde{\mathbf{v}}_n, \ \ \  \mathrm{J}= \left(\begin{array}{ccc}
        \partial_xm_1^x & \partial_ym_1^x & \partial_zm_1^x \\
        \partial_xm_1^y & \partial_ym_1^y & \partial_zm_1^y \\
        \vdots & \vdots & \vdots \\
        \partial_xm_N^z & \partial_ym_N^z & \partial_zm_N^z
    \end{array} \right)
\end{equation}
In this way, Fig.~\ref{fig:translation}(c-d) shows the two pairwise degenerate eigenvectors, which are associated with translation. Both describe the movement of the skyrmion in orthogonal directions.

To analyze this behavior in more detail, the MEP of the translation of skyrmions along the corresponding eigenvectors was calculated for $B_z=1.8$~T for hexagonal lattice models (see Fig.~\ref{fig:translation}(e)) and for the honeycomb lattice (see Fig.~\ref{fig:translation}(f)) using the GNEB method. The energy barriers for the translation of a skyrmion in the honeycomb lattice are approximately a factor of $1000$ larger than for the translation of a skyrmion in the hexagonal lattice. 
This is a strong fingerprint of lattice effects arising from different geometries and thus most probably has to be related to differences in geometric frustration.

\begin{table*}[t]
    \caption{Magnetic interaction constants obtained from DFT. Shell-resolved Heisenberg exchange constants ($J_i$) and DMI constants ($D_i$) were obtained by fitting the energy contribution to spin spirals without and with SOC from DFT calculations as presented in Fig. \ref{spin-spiral}. $J_i>0$ ($J_i<0$) corresponds to FM (AFM) exchange, $D_i>0$ ($D_i<0$) favors CW (CCW) cycloidal spirals, and $K>0~(K<0)$ stabilizes out-of-plane (in-plane) magnetization. A collective hexagonal lattice was used to fit the FGT and the CGT layers. For the CGT layer, the interaction constants on a honeycomb lattice can be directly calculated using the given weighting factors.} \label{table_dmi} 
	\centering
	\scalebox{0.88}{
		\begin{tabular}{ccccccccccccccccccc}
			\hline\hline
Layer & Lattice & $J_1$ & $J_2$ & $J_3$ & $J_4$ & $J_5$ & $J_6$ & $J_7$ & $J_8$ & $D_1$ & $D_2$ & $D_3$ & $D_4$ & $D_5$ & $D_6$ & $D_7$ & $K$\\ 
\hline
FGT& hexagonal (meV/uc) & 59.32  & -2.18 & -3.06 & 0.39 & 3.02 & -1.19 & -0.06 & -0.39 & -0.36  & -0.19 & -0.01 & 0.03 & 0.03 & -0.005 & -0.03 & 9.81\\
Interlayer & hexagonal (meV/uc) & 0.25 & -1.10 & 2.27 & 0.91 & -3.16 & -1.94 & 2.08 & 1.29 & 0.07  & -0.07 & 0.004 & -0.22 & 0.34 & 0.08 & 0.009 & -- \\ 
CGT& hexagonal (meV/uc) & 10.23 & 1.88 & 2.00 & -0.23 & -0.35 & -0.13 & -0.24 & 0.05 & 0.125 & -0.013 & 0.003 & 0.013 & -0.010 & 0.012 & -0.009 & -0.26 \\
CGT& weighting factor ($w_i$) & 2 & 1 & 2 & 2 & 1 & 1 & 2 & 2 & 2 & 1 & 2 & 2 & 1 & 1 & 2 & 1 \\
CGT& honeycomb (meV/atom) & 20.45 & 1.88 & 4.0 & -0.45 & -0.35 & -0.13 & -0.48 & 0.1 & 0.249 & -0.013 & 0.007 & 0.025 & -0.01 & 0.012 & -0.019 & -0.13 \\
			\hline
	\end{tabular}}
\end{table*}	

Nevertheless, for both lattices, skyrmion translations cost energies on the order of $\mu$eV, which are negligible compared to skyrmion energy barriers (on the order of meV).
The harmonic approximation of the energy along the translation, described by eigenvector $\Tilde{\mathbf{v}}=\Tilde{\mathbf{v}}_1,\Tilde{\mathbf{v}}_2$ with coordinates $t=t_1,t_2$
\begin{equation}
    E(\mathbf{m} + \Tilde{\mathbf{v}}t) \approx E(\mathbf{m}) + \frac{1}{2}\lambda t^2
\end{equation}
predicted by the corresponding eigenvalue from Fig.~\ref{fig:translation}(a) and (b) agrees well with 
the MEP in the vicinity of the minimum. The eigenvalues in Fig.~\ref{fig:translation}(b) also predict that the pinning in of the honeycomb lattice skyrmions increases with increasing $B_z$. This can be explained well by the energy contribution along the MEPs in Fig.~\ref{fig:translation}(f). For the skyrmion in the honeycomb lattice, the main contribution to the energy barrier of the translation comes from the Zeeman energy $E_\text{zee}$, while the skyrmion translation in the hexagonal lattice is favored by the Zeeman energy. Increasing the field $B_z$ also increases the Zeeman energy and thus also makes the pinning of the skyrmion on the honeycomb lattice stronger.

\section{CONCLUSIONS AND OUTLOOK}
\label{conclusions}

In summary, by employing an atomistic spin model parameterized from first-principles, we predict the emergence of multiple topological magnetic textures in the all-magnetic vdW heterostructure FGT/CGT. We present an efficient spin-spiral approach for computing magnetic interaction constants from first-principles in hexagonal and honeycomb lattices. We demonstrate this approach in two distinct systems: a hexagonal lattice with a multilayer structure of FGT and a honeycomb lattice of CGT. Topological magnetic solitons with Néel-type helicity are formed at both sides of the FGT/CGT heterostructure with opposite chirality. Skyrmions at the FGT layer and bimerons at the CGT layer persist in the heterostructure at zero field, while the latter undergoes bimeron-skyrmion transformation if a field is applied. 

We provide a detailed discussion of soliton size, energy barriers, collapse mechanism, and the effects of lattice symmetry. In particular, in CGT, we show that although the conversion between hexagonal and honeycomb Hamiltonians leaves the energy of spin spirals as well as the spin stiffness invariant, discrete lattice effects arise that influence the stability and behavior of magnetic solitons. Among these, we identify fundamental differences in terms of geometric frustration and point out the inequality of spin spiral decompositions of solitons in both lattice geometries. As a consequence of these discretization effects, the energy barrier that protects the soliton from the collapse into the field-polarized state can vary by a factor of 2 for solitons of the same size but on different lattices. Finally, we find that skyrmions are more strongly pinned in the honeycomb lattice than in the hexagonal lattice.

\begin{figure}[b]
    \centering    \includegraphics[width=1.0\linewidth]{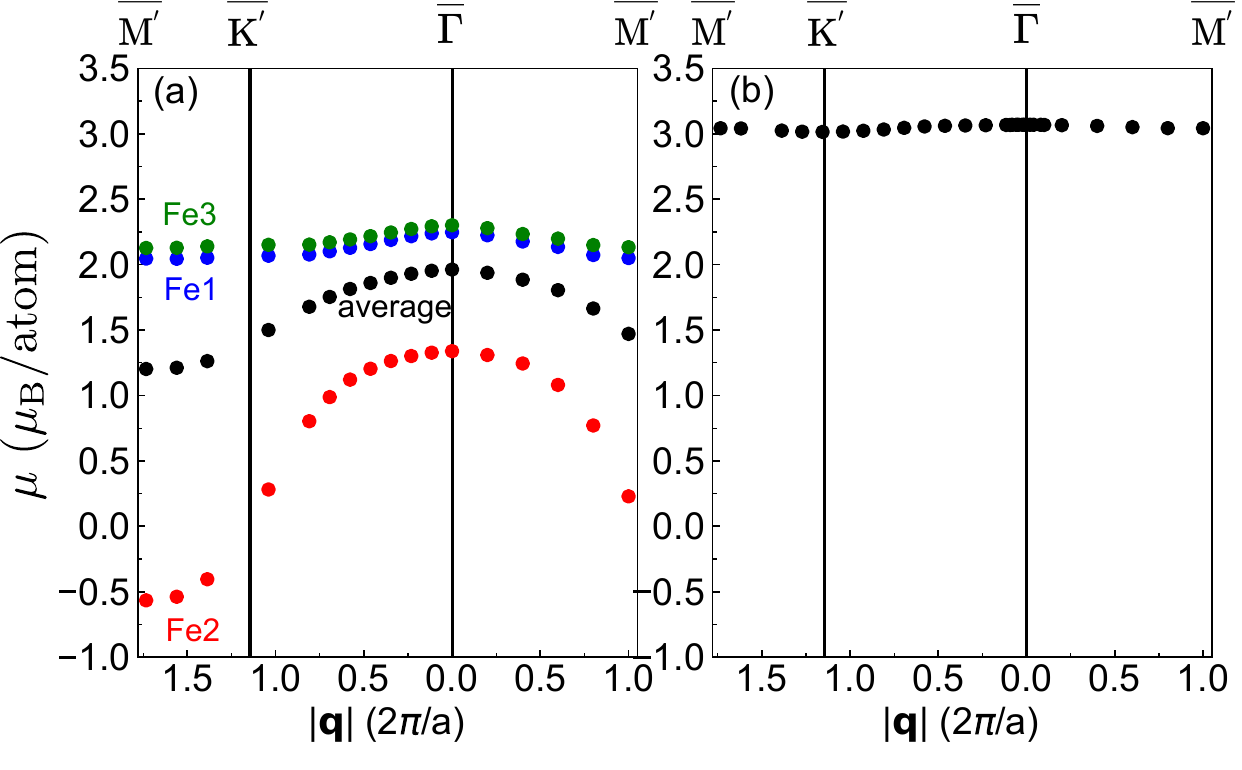}
    \caption{
Variation of magnetic moments obtained via DFT for spin spirals (cf.~Fig.~\ref{spin-spiral})
along the high-symmetry direction ${\overline{\text{M}'\text{K}'\Gamma\text{M}'}}$ in the extended BZ zone for (a) the FGT and (b) the CGT layer. Note that for both layers the average spin moment per magnetic atom, i.e.~Fe or Cr, is given.} 
    \label{fig:spin_moment}
\end{figure}

For future applications, all-magnetic vdW heterostructures represent a promising platform for realizing multiple forms of topological magnetism within a single system, with their stability readily tunable by external stimuli such as gating, strain, or twisting \cite{burch2018magnetism,Dongzhe_prb2023}. Moreover, all-magnetic vdW heterostructures could also be good candidates for synthetic antiferromagnetic skyrmions \cite{pham2024fast}, which could lead to the fast motion of skyrmions by current with the vanishing skyrmion Hall effect \cite{juge2022skyrmions}. Our work represents a step forward in the understanding of the fundamental properties of topological spin textures in 2D vdW magnets and can help to pave the way toward skyrmionic devices based on atomically thin vdW heterostructures.

\section*{ACKNOWLEDGEMENTS} This study has been supported through ANR Grant No. ANR-22-CE24-0019. This work is supported by France 2030 government investment plan managed by the French National Research Agency under grant reference PEPR SPIN – [SPINTHEORY] ANR-22-EXSP-0009. We gratefully acknowledge financial support from the Deutsche Forschungs Gemeinschaft (DFG, German Research Foundation) through SPP2137 “Skyrmionics” (Project No. 462602351). H.S. acknowledges financial support from the Icelandic Research Fund (Grant No. 239435). This work was performed using HPC resources from CALMIP (Grant 2022/2025-[P21023]) and HPC resources available at the Kiel University Computing Centre. 

\section*{DATA AVAILABILITY}
The data that support the findings of this article are not publicly available. The data are available from the authors upon reasonable request.

\section*{APPENDIX A: MAGNETIC INTERACTION PARAMETERS FOR BOTH HEXAGONAL AND HONEYCOMB LATTICES}

Table \ref{table_dmi} lists all relevant magnetic interactions in the hexagonal and honeycomb lattices, along with the relations between them, including the corresponding weighting factors used to map the hexagonal lattice onto the honeycomb lattice for the CGT layer.

Here, we discuss the physical meanings of these interactions in different lattices.
Since the parameters are obtained for collective models of each CGT and FGT, it is important to mention which specific interactions between atoms they contain. For the CGT layer, the honeycomb lattice model is able to contain the full set of Heisenberg interactions and DMI between the magnetic Cr atoms. Each atom on the model lattice corresponds to one Cr atom. Interactions that depend on the small induced magnetization of the Ge and Te atoms are mapped effectively into the interactions between the Cr atoms. 

For the FGT layer, the large number of 9 atoms per unit cell makes it impossible to associate an interaction parameter with specific interacting atoms. Instead, interactions between and inside the individual Fe layers in the FGT are mapped onto the same effective parameters based on their distance inside the plane of the film. Since all spin spiral calculations were done starting from a FM ground state of the FGT, the model assumes a FM alignment between the Fe layers. While this excludes certain antiferromagnetic states of high energy, such as the layer-wise antiferromagnet, it retains all information needed to describe magnetic configurations near the FM ground state, such as skyrmions, and even thermodynamic properties, as demonstrated previously in 
Ref.~\cite{Dongzhe_PRB2024}.

\begin{figure}[t]
    \centering
    \includegraphics[width=1.0\linewidth]{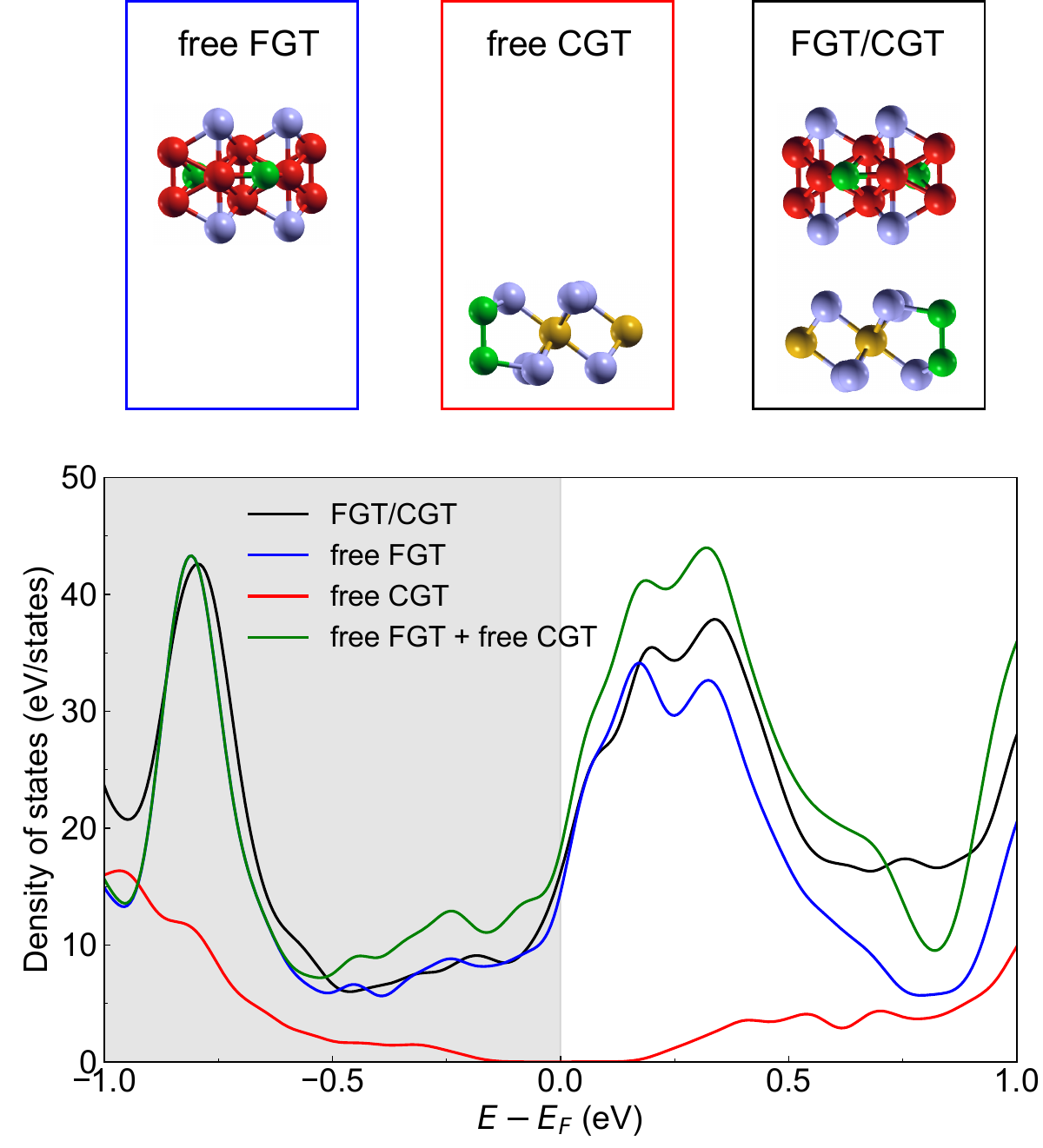}
    \caption{Hybridization effect at the FGT/CGT interface. 
The total density of states (sum of spin-up and -down channels) obtained via DFT is given for freestanding FGT,
but structurally deformed as at the FGT/CGT interface (blue), free-standing CGT, deformed as at the FGT/CGT interface (red), the sum of the free-standing, deformed FGT and CGT (green), and the FGT/CGT interface (black). Note that this hybridization effect is taken into account in the DFT calculations of magnetic interactions (cf.~Fig.~\ref{spin-spiral}).}
    \label{fig:dos}
\end{figure}

\section*{APPENDIX B: SPIN MOMENT VARIATION IN RECIPROCAL SPACE}

In the spin-spiral approach employed in this work to map the DFT total energies, we assume that the spin moments do not change with respect to $\V{q}$. However, as shown in Fig.~\ref{fig:spin_moment}(a), we find from our DFT calculations that the spin moment varies in the FGT layer if one considers the entire range of spin spirals, as also observed in other FGT heterostructures \cite{Dongzhe2022_fgt}. However, in a considerable range around the $\overline{\Gamma}$ point corresponding to the FM state, the magnetic moment varies little. In contrast, for the CGT layer, it remains nearly constant as shown in Fig.~\ref{fig:spin_moment}(b), indicating the strong validity of the spin-spiral approach for the CGT layer.

\section*{APPENDIX C: HYBRIDIZATION EFFECT AT THE INTERFACE}

To quantify only the degree of the hybridization effect at the FGT/CGT interface, we carried out two artificial DFT calculations to obtain the density of states (DOS) of free-standing FGT and CGT, each deformed according to the atomic relaxation at the FGT/CGT interface (see Fig.~\ref{fig:dos}). In this way, we include the effect of structural relaxation.

We find that FGT exhibits a much larger DOS compared to CGT. By summing the contributions of the two free-standing layers, we obtain the total DOS without accounting for the hybridization effect (plotted as green). When comparing the blue curve with the black one, which includes the hybridization effect, we find that the qualitative behavior remains nearly the same, but there is a quantitative discrepancy of about 10\%. We also note that this discrepancy appears relatively large when compared to the DOS of free-CGT. However, when comparing the green and black curves, the total difference remains on the order of 12\% (with the qualitative behavior unchanged), which is consistent with the relatively weak van der Waals interaction between the FGT and CGT layers. Nevertheless, in our DFT calculations presented in the main text, we have included the hybridization effect.

\bibliography{References}

\end{document}